\documentstyle[emulateapj,apjfonts,psfig]{article}

\slugcomment{Accepted for publication by {\it The Astrophysical Journal Letters}
July 11, 2001}

\lefthead{Freire et al.}
\righthead{Ionized Gas in 47~Tuc}

\begin{document}

\title{Detection of Ionized Gas in the Globular Cluster 47~Tucanae}

\author{P.~C.~Freire, M.~Kramer, and A.~G.~Lyne}
\affil{The University of Manchester, Jodrell Bank Observatory,
  Macclesfield, Cheshire, SK11 9DL, UK}
\author{F.~Camilo}
\affil{Columbia Astrophysics Laboratory, Columbia University,
  550~West~120th Street, New York, NY~10027}
\author{R.~N.~Manchester}
\affil{Australia Telescope National Facility, CSIRO, P.O.~Box~76,
  Epping, NSW~1710, Australia}
\and
\author{N.~D'Amico}
\affil{Osservatorio Astronomico di Bologna, via Ranzani 1,
  40127~Bologna, Italy}

\begin{abstract}
We report the detection of ionized intracluster gas in the globular
cluster 47~Tucanae.  Pulsars in this cluster with a negative
period derivative, which must lie in the distant half of the cluster,
have significantly higher measured integrated electron column densities
than the pulsars with a positive period derivative.  We derive the
plasma density within the central few pc of the cluster using two
different methods which yield consistent values.  Our best estimate of
$n_e = 0.067\pm0.015$\,cm$^{-3}$ is about 100 times the free electron
density of the ISM in the vicinity of 47~Tucanae, and the ionized gas
is probably the dominant component of the intracluster medium.
\end{abstract}


\section{Introduction} \label{sec:intro}

Every $\sim10^8$\,yr, globular clusters pass through the plane of the
Galaxy, during which any intracluster gas is expected to be stripped
from the systems (\cite{spe91}).  However, winds from evolved stars
continuously fill the cluster with gas between passages.  Heretofore,
numerous searches for the expected neutral and ionized material have
been unsuccessful (see references in \cite{spe91}; \cite{hs77};
\cite{swfw90}; \cite{kgbv96}; \cite{peo97}; \cite{hee+99}) or
inconclusive (\cite{kg95}; \cite{osa+97}). The most convincing evidence
presented so far was a 3\,$\sigma$ detection of two CO lines in the
direction of 47~Tucanae (henceforth 47~Tuc). These were interpreted as
resulting from a bow shock formed as the cluster moves in the Galactic
halo (\cite{ogff97}), indicating the presence of an intracluster medium.
There is also an unconfirmed 3.5\,$\sigma$ SCUBA detection of dust in
NGC~6356 at $\lambda = 850$\,nm (\cite{hepe98}), and a tentative
detection of diffuse X-ray emission by the intracluster medium of
NGC~6779 (\cite{her+00}).

In this Letter we report the detection of ionized gas in 47~Tuc from
new measurements of the radio dispersion of 15 of the 20 millisecond
pulsars known in the cluster (\cite{clf+00}).

\section{Observations} \label{sec:obs}

Observations of 47~Tuc using the Parkes radio telescope at a frequency
of 1.4\,GHz have been used to measure the precise positions, periods
$P$, and apparent rates of change of period $\dot P_{\rm obs}$, for 15
of its 20 known pulsars (\cite{fcl+01}).  We have continued timing
observations of the pulsars in this cluster, and since 1999 August have
used a new system providing a three-fold increase in time resolution.
This consists of the central beam of the Parkes telescope multibeam
receiver, as before, but uses a $2\times512\times0.5$-MHz filter bank
centered on 1390\,MHz, and a sampling interval of $80\,\mu$s.  With
these data we acquire very high quality pulse profiles, from which we
obtain the precise dispersion measure for each pulsar (DM, the
integrated electron column density along the line of sight from the
Earth), from the relative delays in the arrival times of pulses in four
contiguous 64-MHz subbands.

We do this with the {\sc tempo} timing
software\footnote{http://pulsar.princeton.edu/tempo} as follows.  First
we determine timing solutions for all pulsars including spin,
astrometric, and binary parameters where relevant, as detailed in
Freire et al.~(2001\nocite{fcl+01}), but using only data collected at
1.4\,GHz (medium-resolution data during 1997 August--1999 August and
high-resolution data during 1999 August--2001 February).  In virtually
all cases the parameters obtained are consistent at the 3\,$\sigma$
level with those reported by Freire et al.~(2001\nocite{fcl+01}).  We
then hold the pulsar ephemerides fixed at these values, and fit for DM,
using only high-resolution times-of-arrival measured in four frequency
subbands.  The resulting DMs are listed in Table~\ref{tab:parms}, along
with the $(\dot{P}/P)_{\rm obs}$, and angular offset from the center of
the cluster for each pulsar, $\theta_\perp$, as determined by Freire et
al.~(2001\nocite{fcl+01}).

As Figure~\ref{fig:47tuc} shows, all 15 pulsars are located near the
center of the cluster.  Having periods of a few milliseconds, they are
expected to have the small positive intrinsic period derivatives
$\dot{P}_{\rm int}$ which are typical of millisecond pulsars.  However,
the gravitational field of the cluster causes the pulsars to accelerate
toward its center.  The line-of-sight component $a$ of this
acceleration results in a contribution to the observed period
derivative of each pulsar given by $\dot{P}/P=a/c$, where $c$ is the
speed of light.  In the central regions of 47~Tuc, these values are
typically greater than the ${(\dot{P}/P)}_{\rm int}$ for millisecond
pulsars (\cite{fcl+01}), so that the pulsars act as tracers of the
cluster gravitational field.

\medskip
\centerline{ \psfig{file=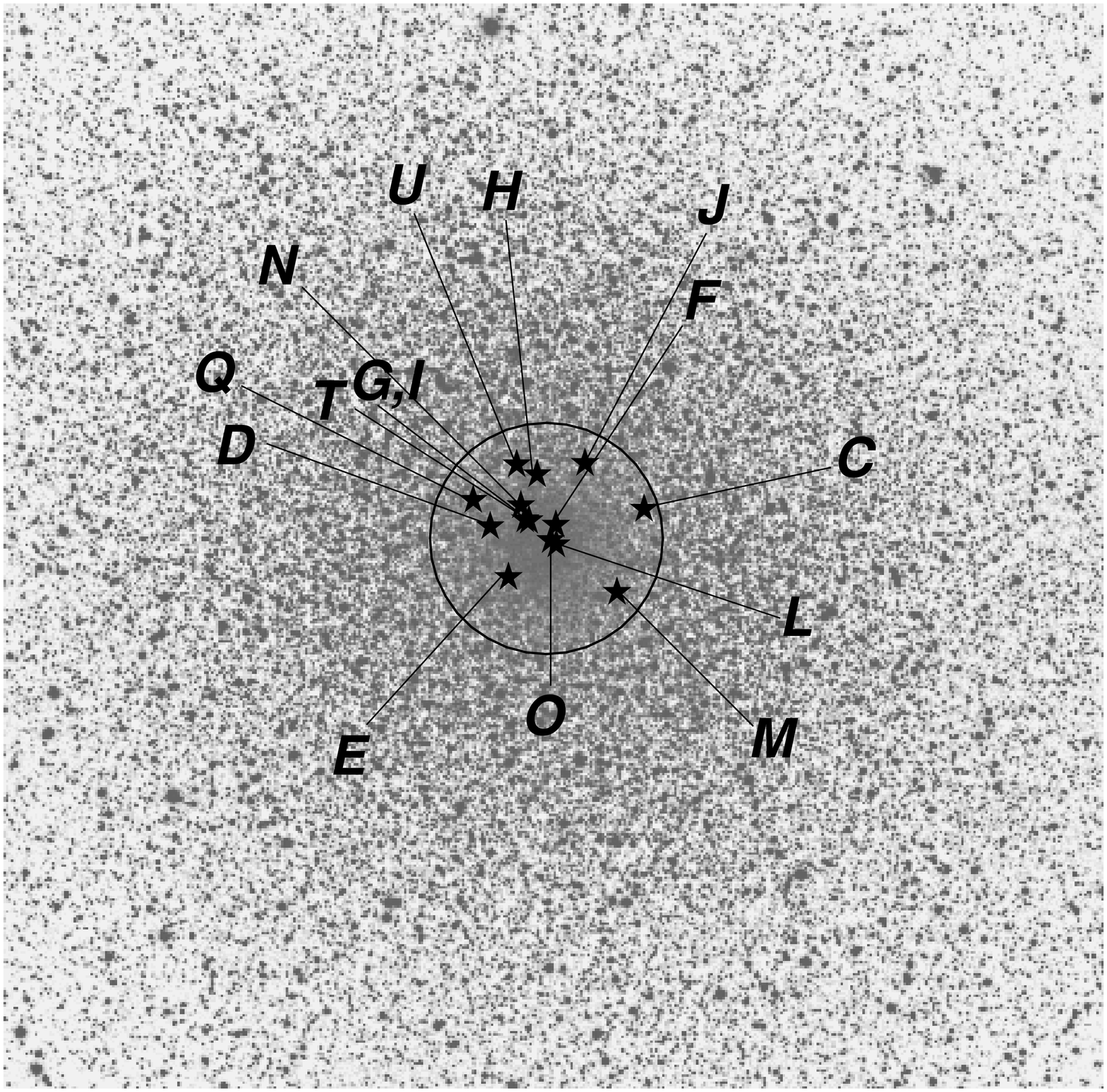,height=8cm,angle=0} }
\medskip
\figcaption[figure1.ps]{\label{fig:47tuc} 
The positions of 15 millisecond pulsars in 47~Tuc superimposed on an
optical image of the cluster (obtained by D.~Malin, Anglo-Australian
Observatory).  All pulsars are located within $1\farcm2$ ($\approx
2$\,pc, delineated by the circle) of the center of the cluster
(\cite{fcl+01}).  North is to the top and east is to the left. }
\bigskip

Conversely, from a knowledge of the gravitational field of the globular
cluster, one can use an observed period derivative to constrain the
position of a pulsar in the cluster along the line of sight.  In
particular, pulsars located on the far side of the cluster will
experience an acceleration directed towards the Earth, resulting in a
negative period derivative which often exceeds the intrinsic value,
leading to a negative value of $(\dot{P}/P)_{\rm obs}$.  We observe
this for nine of the 15 pulsars (Table~\ref{tab:parms}).
Figure~\ref{fig:DMacc} shows the measured DMs plotted against
$(\dot{P}/P)_{\rm obs}$.  Strikingly, most of the nine pulsars with
negative $(\dot{P}/P)_{\rm obs}$, which {\em must\/} lie in the distant
half of the cluster, have significantly higher DMs than those with
positive $(\dot{P}/P)_{\rm obs}$, most of which are likely to reside on
the near side.

We now show that this configuration is unlikely to arise by chance, by
computing the chance probability of observing (as we do) the 7 pulsars
with the highest DMs all having a negative $\dot{P}$ (we neglect in
this calculation 47~Tuc L and T, whose DMs and uncertainties make their
contribution to this argument uncertain).  The number of sets of 7
pulsars among the 13 with a timing solution is $13!/(7! \times 6!) =
1716$. The number of sets of 7 pulsars among the 8 with a negative
$\dot{P}$ is 8.  Therefore the probability that the observed
arrangement is due to chance is small, $8/1716 \approx 0.005$.  It is
conceivable that the observed variations in DM are simply a result of
electron density variations along the different lines of sight in the
interstellar medium.  However this is very unlikely, because the
variations in DM caused by irregularities in the Galactic electron
column density are at the level $\delta {\rm DM} \sim
0.05$\,cm$^{-3}$\,pc over the small angle subtended by the pulsars
(\cite{nct92}), while the variations we observe are 10 times larger
than this.

We conclude that the most likely cause of the observed DM variations is
the presence of a free electron plasma within the cluster, and
estimate its density in the next section.

\medskip
\centerline{ \psfig{file=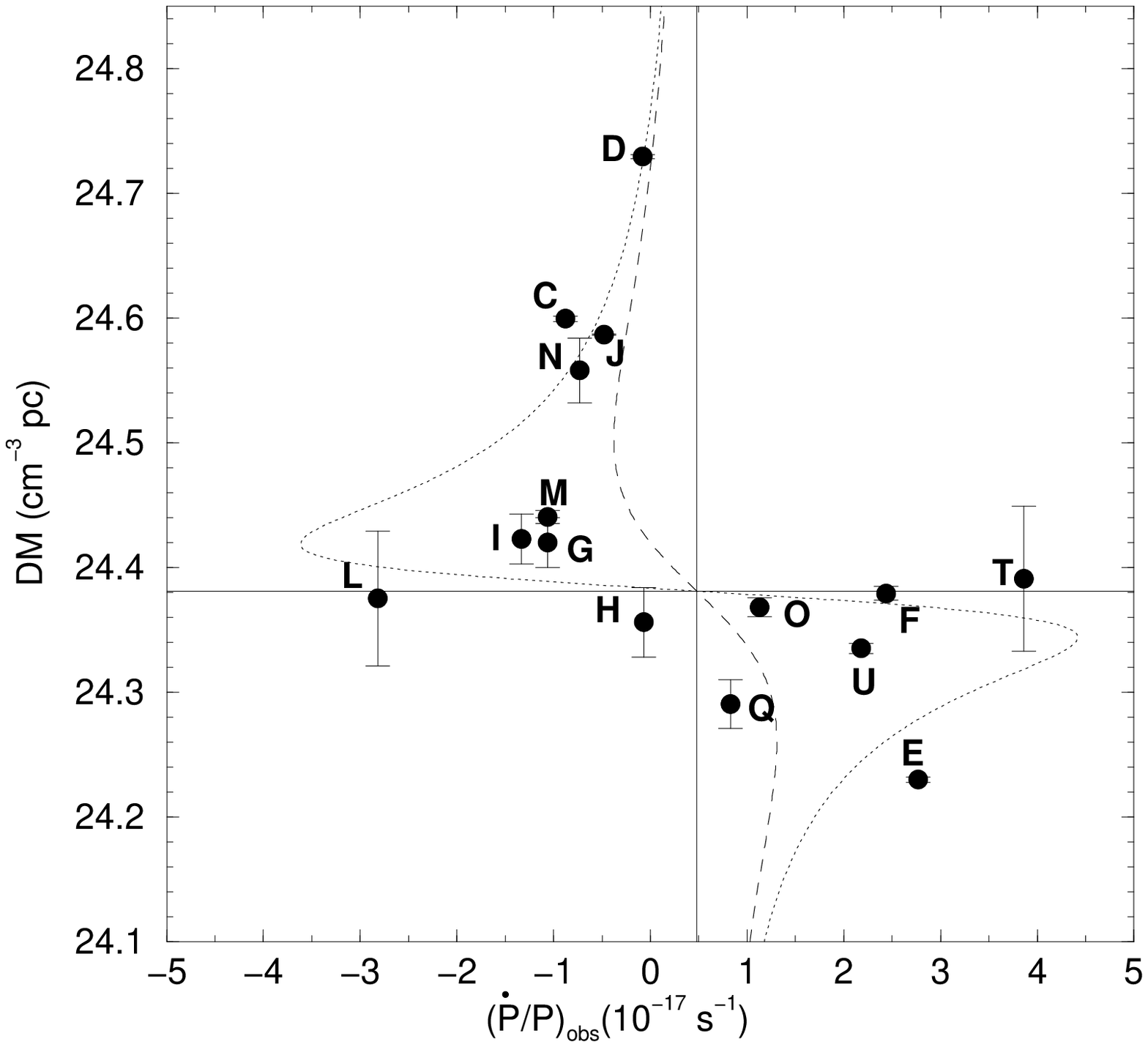,height=8cm,angle=0} }
\medskip
\figcaption[figure2.ps]{\label{fig:DMacc} Measured DM plotted
against ${(\dot{P}/P)}_{\rm obs}$ for 15 millisecond pulsars.  Most of
the pulsars with negative $(\dot{P}/P)_{\rm obs}$ have significantly
higher DMs than those with positive $(\dot{P}/P)_{\rm obs}$.  The
curves show the expected observed variation of ${(\dot{P}/P)}$ with DM
along lines of sight passing through the cluster core ({\em dotted\/})
and passing 2\,pc from the cluster core ({\em dashed\/}).  These are
calculated using the cluster mass distribution described in Freire et
al.~(2001)\protect\nocite{fcl+01}, assuming that all the pulsars have
the same intrinsic $\langle(\dot{P}/P)_{\rm int}\rangle$, and that
there is a uniform electron density of 0.067\,cm$^{-3}$ throughout the
cluster.  The vertical and horizontal lines represent the fitted values
of $\langle(\dot{P}/P)_{\rm int}\rangle$ and DM$_{\rm c}$ as described
in \S~\ref{sec:ne}. }
\bigskip

\section{Plasma Density} \label{sec:ne}

We derive the plasma density in the central regions of 47~Tuc using two
different methods.  In the first, and simpler method, the density $n_e$ of a
uniform plasma is obtained from the ratio of dispersion in measured DMs
to the one-dimensional dispersion in offsets in the plane of the sky
from the cluster center (Table~\ref{tab:parms}):
\begin{equation}
n_e \simeq \frac{\sigma \mbox{DM}}{\sigma (\theta_\perp D)} =
\frac{0.13\pm0.04\,\mbox{cm}^{-3}\,\mbox{pc}}{0.75\pm0.14\,\mbox{pc}} =
0.17\pm0.05\,\mbox{cm}^{-3},
\label{eq:ne1}
\end{equation}
where we have used $D=5$\,kpc (\cite{rei98b}), and assume a spherically
symmetric distribution of pulsars.

In the second method we calculate the positions of the pulsars along
the line of sight from their period derivatives and compare these with
the DMs. The measured values of $\dot{P}/P$ are biased by the intrinsic
period derivative of each pulsar: with no a priori knowledge of
$(\dot{P}/P)_{\rm int}$ for individual pulsars, we assume that all have
the average value $\langle(\dot{P}/P)_{\rm int}\rangle$.  Then,
from the inferred accelerations $a/c = (\dot{P}/P)_{\rm obs} -
\langle(\dot{P}/P)_{\rm int}\rangle$ and a King model for the
gravitational potential of the cluster (\cite{kin66}; see
\cite{fcl+01}), we compute the distance $R$ of each pulsar along the
line of sight from the plane of the sky containing the center of the
cluster.  Here we are assuming that a King model for the potential
computed with known cluster parameters, together with a relatively
small contribution from $(\dot{P}/P)_{\rm int}$, provides a good
description for the observed values of $\dot{P}/P$, which Freire et
al.~(2001\nocite{fcl+01}) have shown to be the case.  Using a uniform
electron density $n_e$, we then calculate the incremental DM with
respect to the DM to the center of the cluster, DM$_{\rm c}$, required
to obtain the observed DM for each pulsar.  In summary, our model for
the DMs contains the free parameters DM$_{\rm c}$, $n_e$, and
$\langle{(\dot{P}/P)}_{\rm int}\rangle$, and is represented by
\begin{equation}
{\cal DM}_i = {\rm DM_c} + n_e R_i ([\dot{P}/P]_{{\rm obs}\,i}
- \langle[\dot{P}/P]_{\rm int}\rangle),
\label{eq:ne2a}
\end{equation}
where we have noted the explicit dependence of each pulsar's
line-of-sight distance $R_i$ on the inferred acceleration.  There are
often two distances at which a given $a/c$ can be obtained (see
Fig.~\ref{fig:DMacc}).  We resolve this ambiguity by choosing the
distance that produces a model dispersion measure ${\cal DM}$ closest
to the observed value DM.

We determine the model parameters by minimizing
\begin{equation}
\mu \equiv \sum_{i=1}^{15} ({\rm DM}_i - {\cal DM}_i)^2,
\label{eq:ne2b}
\end{equation}
and in order to obtain reliable uncertainty estimates in the derived
parameters we implement this with a Monte Carlo procedure.  We do so by
generating 10,000 data sets where both the observed pulsar DMs and
relevant cluster parameters are chosen in a random fashion consistent
with their measured uncertainties (distance: $5.0\pm0.4$\,kpc;
\cite{rei98b}; central stellar line-of-sight velocity dispersion:
$11.6\pm1.4$\,km\,s$^{-1}$; \cite{mm86b}; angular core radius:
$23\farcs1\pm1\farcs4$; \cite{hgg00}).

Minimization of $\mu$ (eq.~[\ref{eq:ne2b}]) in the Monte Carlo
procedure resulted in an average and rms for the model parameters\footnote{In general the
parameters obtained in this procedure are not Gaussian distributed.
Where the difference is small we approximate to average and rms values,
and otherwise present the median and one-sided ``34\%'' error bars.}
of $n_e = 0.067\pm0.015$\,cm$^{-3}$, DM$_{\rm
c} = 24.381\pm0.009$\,cm$^{-3}$\,pc and $\langle{(\dot{P}/P)}_{\rm
int}\rangle = (0.48\pm0.18)\times10^{-17}$\,s$^{-1}$.  The latter
corresponds to an average ``characteristic age'' of
$\langle{(P/2\dot{P})}_{\rm int}\rangle = 3$\,Gyr and an average
magnetic field of $\approx 3\times10^8$ Gauss for these 15 millisecond
pulsars in 47~Tuc.  The line-of-sight distances $R_i$ and uncertainties
resulting from the fitting procedure are displayed in
Figure~\ref{fig:DMrad} and listed in Table~\ref{tab:parms}.

Each pulsar's ${(\dot{P}/P)}_{\rm int}$ is different from the average.
In order to investigate the effect of this approximation in the
evaluation of our model parameters we repeated the Monte Carlo
procedure described above, replacing $\langle{(\dot{P}/P)}_{\rm
int}\rangle$ with a value of ${(\dot{P}/P)}_{\rm int}$ for each pulsar
drawn from a Gaussian distribution of characteristic ages centered on
5\,Gyr with an rms of 2.5\,Gyr, truncated at the low end by the limits
derived by Freire et al.~(2001)\nocite{fcl+01}.  We repeated the
procedure for a flat distribution of ages between 2 and 9\,Gyr.  Both
distributions were chosen to be approximately representative of the
distribution of characteristic ages observed for $\sim 30$ millisecond
pulsars in the Galactic disk.  The results were similar in both cases,
yielding parameters smaller by about 1$\,\sigma$ compared to the
original model represented by equation~(\ref{eq:ne2a}):  $n_e =
0.058_{-0.012}^{+0.017}$\,cm$^{-3}$ and DM$_{\rm c} =
24.37\pm0.02$\,cm$^{-3}$\,pc.  Additionally, for three pulsars the
inferred distances $R_i$ or their uncertainties changed significantly,
owing to the prominence of the ``second solution'' in $R_i(a)$ (cf.
Fig.~\ref{fig:DMacc}), and are indicated in Table~\ref{tab:parms}.

Individual DMs are affected by fluctuations in the Galactic column
density, as noted in \S~\ref{sec:obs}.  There is also some uncertainty
regarding the detailed form of the potential used to
derive $R_i$ from the inferred accelerations.  And each pulsar's
characteristic age is different from the average value determined in
the model.  All these factors contribute to the scatter visible in
Figure~\ref{fig:DMrad}.  Nevertheless, the small scatter about the
linear relation between DM and $R$ emphasizes that the observations are
consistent with two of the central assumptions made in the fit:  that
the cloud of ionized gas at the center of 47~Tuc is fairly
homogeneous on $\sim$ pc scales, and that the King model (one simple
choice amid many possible ones; see \cite{phi93}) provides a consistent
description for the potential of the cluster.

\medskip
\centerline{ \psfig{file=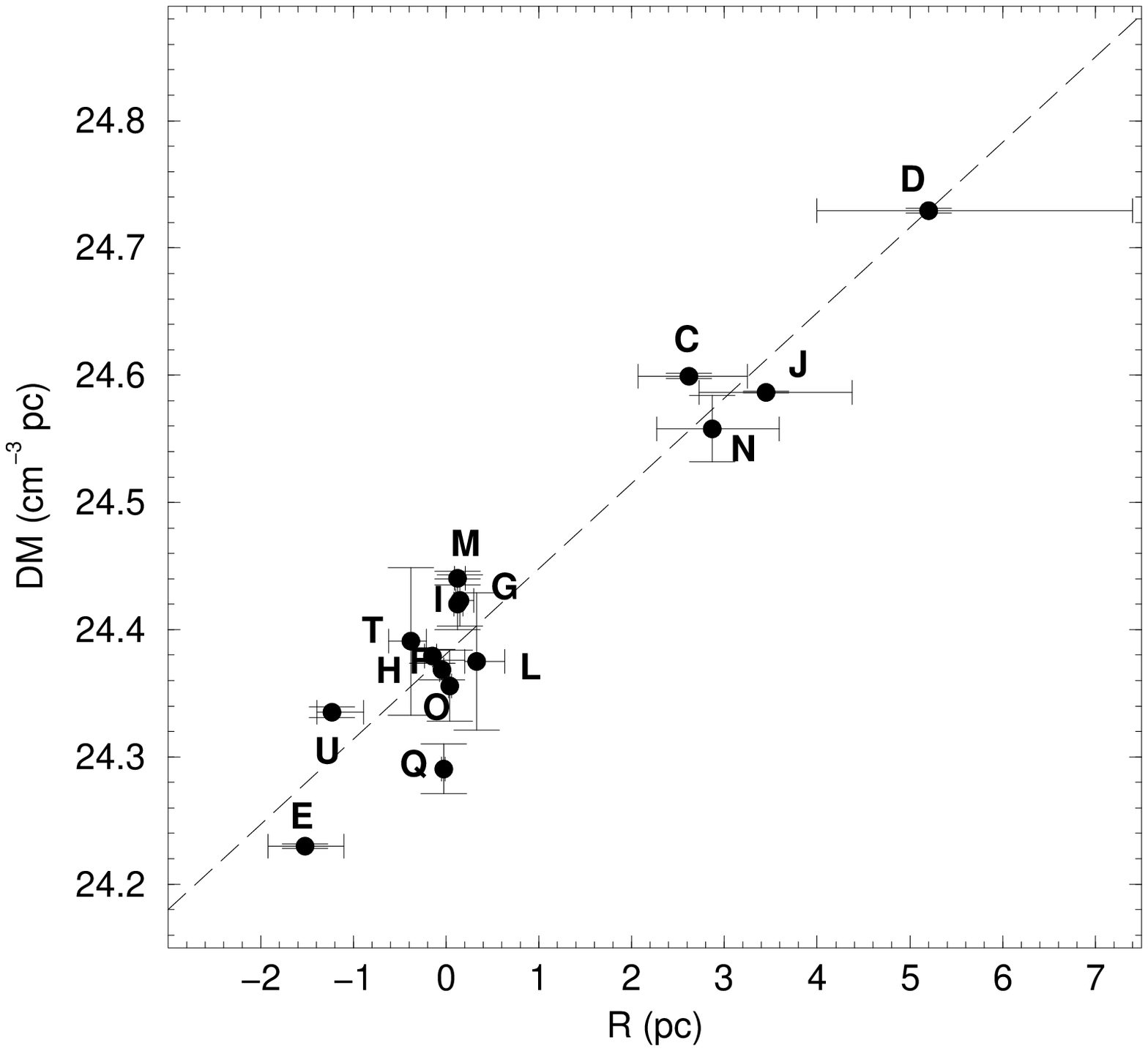,height=8cm,angle=0} }
\medskip
\figcaption[figure3.ps]{\label{fig:DMrad} The measured DM
plotted against the inferred line-of-sight distance of 15 pulsars from
the plane of the sky containing the cluster center.  A strong
correlation is observed, for which the inferred free electron density
is $0.067 \pm 0.015$\,cm$^{-3}$. }
\bigskip

The values of $n_e$ derived by the methods represented by equations
(\ref{eq:ne1}) and (\ref{eq:ne2a}) agree at the $2\,\sigma$ level,
although the value derived in equation~(\ref{eq:ne1}) is nominally
higher than the one derived in the Monte Carlo procedure.  This is
partly due to the apparently smaller effective radius of the pulsar
distribution in the plane of the sky (1.9\,pc; Fig.~\ref{fig:47tuc}),
than along the line of sight (3.3\,pc; Fig.~\ref{fig:DMrad}).  We
attribute this apparent asphericity of the pulsar distribution to
small-number statistics (confirmed with a KS test), and hereafter
consider the pulsars to be distributed in a sphere with effective
radius the geometrical mean of the two, $r_p = 2.5$\,pc.  We also
retain $0.067\pm0.015$\,cm$^{-3}$ as the most precise estimate for
$n_e$.

Eight pulsars with well-measured DM and $(\dot{P}/P)_{\rm obs}$ are
also known in the globular cluster M15. The DM variations among these
pulsars were interpreted by Anderson (1992\nocite{and92}) as arising
from a gradient in the Galactic electron column density across the
cluster.  As an alternative interpretation, we find the data for the
four millisecond pulsars ($P<10$\,ms) in M15 to be consistent with the
existence of a homogeneous plasma with an electron density of about
0.2\,cm$^{-3}$, slightly larger than in 47~Tuc. The other pulsars in
M15 have longer periods and likely have large contaminating values of
$(\dot{P}/P)_{\rm int}$.

\section{Discussion} \label{sec:disc}

Using high precision pulsar timing applied to 15 pulsars observed in
the globular cluster 47~Tuc, we have detected ionized intracluster gas
with a density undetectable by other methods.  Our measurement is
consistent with previous upper limits for 47~Tuc (\cite{hs77}).  The
central electron density derived, $n_e = 0.07$\,cm$^{-3}$, compares to
$\sim 0.0007$\,cm$^{-3}$ for the ISM in the vicinity of 47~Tuc, at a
$z$-height of 3.5\,kpc, and $\sim 0.02$\,cm$^{-3}$ in the solar
neighborhood (\cite{tc93}).

With a value of 0.07\,cm$^{-3}$ for the mean electron density and
assuming that each electron is accompanied by a single proton, we
obtain a plasma mass density $\rho \sim 10^{-25}$\,g\,cm$^{-3}$.
Within the approximately spherical region of radius $r_p=2.5$\,pc
occupied by the pulsars this corresponds to a total mass
$\sim0.1$\,M$_\odot$.  Recent ISO observations (\cite{hee+99}) put a
strong upper limit of $2\times10^{-4}$\,M$_\odot$ on the dust content
in 47~Tuc, so that virtually any intracluster medium should be present
as ionized or neutral gas.  Moreover, we expect any gas present to be
completely ionized: the detection of hot stars in the cluster
(\cite{psm+91}; \cite{gysb92}; \cite{ods+97}) indicates an intense
background of UV photons, enough to ionize at least
$10^{-3.5}$\,M$_\odot$\,yr$^{-1}$.  This is more than enough to
completely ionize the $\dot{M} \sim 10^{-5}$\,M$_\odot$\,yr$^{-1}$
which is lost by all the evolved stars in a typical cluster
(\cite{rob86}). We therefore conclude that the detected plasma
represents the total intracluster medium in the inner few pc.

The inferred $\sim 0.1$\,M$_\odot$ is much less than the $\sim
100$\,M$_\odot$ of intracluster material expected to accumulate within
$r_p$ of the cluster core over a period of $10^7$--$10^8$\,yr
(\cite{rob86}).  What process is then responsible for ejecting most of
the gas from this region?  Several different mechanisms have been
proposed, most of which invoke winds such as those driven by
main-sequence stars (\cite{smi99}), novae (\cite{sd78}), M dwarfs
(\cite{cw77}), or pulsars.  Indeed, Spergel (1991\nocite{spe91})
proposed that 47~Tuc and other globular clusters should be devoid of
any plasma, owing to the kinetic effects of strong winds from
millisecond pulsars. Using the derived $\langle{(\dot{P}/{P})}_{\rm
int}\rangle$, the typical spin-down luminosity ($\propto \dot P/P^3$)
of a millisecond pulsar in 47~Tuc is $\approx 10^{34}$\,erg\,s$^{-1}$.
Thus, the $\sim 10^{34}$\,erg\,s$^{-1}$ needed to expel
$10^{-5}$\,M$_\odot$\,yr$^{-1}$ from the cluster's potential, which
requires an escape velocity of 58\,km\,s$^{-1}$ (\cite{web85}), can be
provided by just $\sim 0.5$\% of the spin-down luminosity of the $\sim
200$ pulsars believed to exist in the cluster (\cite{clf+00}).

If we assume that mass loss within the central few pc of the cluster is
$\dot M \sim f \times 10^{-5}$\,M$_\odot$\,yr$^{-1}$, with $f$ of order
unity (\cite{rob86}), then a steady state situation implies that the
plasma is being expelled from the inner $r_p=2.5$\,pc with velocity $v
= \dot{M}/(4 \pi r_p^2 \rho) \sim f \times 80$\,km\,s$^{-1}$, which is
comparable to the escape velocity of the cluster.  Also, the existence
of this plasma, together with sensitive X-ray limits on accretion
luminosity from a central source, can be used to place an upper limit
on the mass of a central black hole of about 100\,M$_\odot$
(\cite{ghem01}).

To conclude, we have detected the long-sought intracluster medium in a
globular cluster, 47~Tucanae. It is somewhat ironic that the very
objects which allow us to detect the ionized gas might also be
responsible for its low concentration in the cluster.

\acknowledgements

The Parkes radio telescope is part of the Australia Telescope which is
funded by the Commonwealth of Australia for operation as a National
Facility managed by CSIRO.  PCF gratefully acknowledges support from
Funda\c{c}\~{a}o para a Ci\^{e}ncia e a Tecnologia through Praxis~XXI
fellowship number BD/11446/97. FC is supported by NASA grant
NAG~5-9095.  NDA is supported by the Ministero dell'Universit\'{a} e
della Ricerca Scientifica e Tecnologica.


\begin{thebibliography}{}

\bibitem[Anderson 1992]{and92}
Anderson, S.~B. 1992.
\newblock PhD thesis, California Institute of Technology

\bibitem[{Camilo} {et al.}  2000]{clf+00}
{Camilo}, F., {Lorimer}, D.~R., {Freire}, P., {Lyne}, A.~G., \& {Manchester},
  R.~N. 2000, ApJ, 535, 975

\bibitem[{Coleman} \& {Worden} 1977]{cw77}
{Coleman}, G.~D. \& {Worden}, S.~P. 1977, ApJ, 218, 792

\bibitem[{Freire} {et al.}  2001]{fcl+01}
{Freire}, P.~C., {Camilo}, F., {Lorimer}, D.~R., {Lyne}, A.~G., {Manchester},
  R.~N., \& {D'Amico}, N. 2001, MNRAS.
\newblock In press, astro-ph/0103372

\bibitem[Grindlay {et al.}  2001]{ghem01}
Grindlay, J.~E., Heinke, C., Edmonds, P.~D., \& Murray, S.~S. 2001, Science,
\newblock 292, 2290

\bibitem[Guhathakurta {et al.}  1992]{gysb92}
Guhathakurta, P., Yanny, B., Schneider, D.~P., \& Bahcall, J.~N. 1992, AJ, 104,
  1790

\bibitem[{Hesser} \& {Shawl} 1977]{hs77}
{Hesser}, J.~E. \& {Shawl}, S.~J. 1977, ApJ, 217, L143

\bibitem[{Hopwood} {et al.}  1998]{hepe98}
{Hopwood}, M. E.~L., {Evans}, A., {Penny}, A., \& {Eyres}, S. P.~S. 1998,
  MNRAS, 301, L30

\bibitem[{Hopwood} {et al.}  2000]{her+00}
{Hopwood}, M. E.~L. {et al.}  2000, MNRAS, 316, L5

\bibitem[{Hopwood} {et al.}  1999]{hee+99}
{Hopwood}, M. E.~L., {Eyres}, S. P.~S., {Evans}, A., {Penny}, A., \&
  {Odenkirchen}, M. 1999, A\&A, 350, 49

\bibitem[{Howell}, {Guhathakurta}, \& {Gilliland} 2000]{hgg00}
{Howell}, J.~H., {Guhathakurta}, P., \& {Gilliland}, R.~L. 2000, PASP, 112,
  1200

\bibitem[King 1966]{kin66}
King, I. 1966, AJ, 71, 64

\bibitem[{Knapp} {et al.}  1996]{kgbv96}
{Knapp}, G.~R., {Gunn}, J.~E., {Bowers}, P.~F., \& {Vasquez Poritz}, J.~F.
  1996, ApJ, 462, 231

\bibitem[{Krockenberger} \& {Grindlay} 1995]{kg95}
{Krockenberger}, M. \& {Grindlay}, J.~E. 1995, ApJ, 451, 200

\bibitem[{Meylan} \& {Mayor} 1986]{mm86b}
{Meylan}, G. \& {Mayor}, M. 1986, A\&A, 166, 122

\bibitem[{Nordgren}, {Cordes}, \& {Terzian} 1992]{nct92}
{Nordgren}, T.~E., {Cordes}, J.~M., \& {Terzian}, Y. 1992, AJ, 104, 1465

\bibitem[{O'Connell} {et al.}  1997]{ods+97}
{O'Connell}, R.~W. {et al.}  1997, AJ, 114, 1982

\bibitem[{Origlia} {et al.}  1997a]{ogff97}
{Origlia}, L., {Gredel}, R., {Ferraro}, F.~R., \& {Fusi Pecci}, F. 1997a,
  MNRAS, 289, 948

\bibitem[{Origlia} {et al.}  1997b]{osa+97}
{Origlia}, L., {Scaltriti}, F., {Anderlucci}, E., {Ferraro}, F.~R., \& {Fusi
  Pecci}, F. 1997b, MNRAS, 292, 753

\bibitem[Paresce {et al.}  1991]{psm+91}
Paresce, F., Shara, M., Meylan, G., Baxter, D., \& Greenfield, P. 1991, Nature,
  352, 297

\bibitem[{Penny}, {Evans}, \& {Odenkirchen} 1997]{peo97}
{Penny}, A.~J., {Evans}, A., \& {Odenkirchen}, M. 1997, A\&A, 317, 694

\bibitem[Phinney 1993]{phi93}
Phinney, E.~S. 1993, in { Structure and Dynamics of Globular Clusters}, ed.\
  S.~G. Djorgovski \& G. Meylan, Astronomical Society of the Pacific Conference
  Series, 141

\bibitem[Reid 1998]{rei98b}
Reid, N. 1998, AJ, 115, 204

\bibitem[{Roberts} 1986]{rob86}
{Roberts}, M. 1986, in { IAU Symp.~126: Harlow Shapely Symposium on Globular
  Cluster Systems in Galaxies}, ed.\ J.F. Grindlay \& A.G.D Philip, (Dordrecht:
  Kluwer), 411

\bibitem[{Scott} \& {Durisen} 1978]{sd78}
{Scott}, E.~H. \& {Durisen}, R.~H. 1978, ApJ, 222, 612

\bibitem[{Smith} 1999]{smi99}
{Smith}, G.~H. 1999, PASP, 111, 980

\bibitem[{Smith} {et al.}  1990]{swfw90}
{Smith}, G.~H., {Wood}, P.~R., {Faulkner}, D.~J., \& {Wright}, A.~E. 1990, ApJ,
  353, 168

\bibitem[Spergel 1991]{spe91}
Spergel, D.~N. 1991, Nature, 352, 221

\bibitem[Taylor \& Cordes 1993]{tc93}
Taylor, J.~H. \& Cordes, J.~M. 1993, ApJ, 411, 674

\bibitem[Webbink 1985]{web85}
Webbink, R.~F. 1985, in { Dynamics of Star Clusters, {IAU} {S}ymposium {N}o.
  113}, ed.\ J. Goodman \& P. Hut, (Dordrecht: Reidel), 541

\end{thebibliography}


\begin{deluxetable}{cccll}
\tablecaption{\label{tab:parms}Parameters for 15 millisecond
pulsars in 47~Tuc }
\tablecolumns{5}
\tablewidth{0pc}
\tablehead{
\colhead{Pulsar}                  &
\colhead{$(\dot{P}/P)_{\rm obs}$} &
\colhead{$\theta_{\perp}$}        &
\colhead{DM}                      &
\colhead{$R$\tablenotemark{a}}  \nl
\colhead{}                        & 
\colhead{$(10^{-17}\rm s^{-1})$}  &
\colhead{(arcmin)}                &
\colhead{(cm$^{-3}$\,pc)}         &
\colhead{(pc)}                 \nl}
\startdata
 C & $-$0.91 & 1.21 & 24.599(2) & $+2.6\pm0.6$            \nl
 D & $-$0.10 & 0.68 & 24.729(2) & $+5.2^{+2.2}_{-1.2}$    \nl
 E & $+$2.74 & 0.65 & 24.230(2) & $-1.5\pm0.4$            \nl
 F & $+$2.42 & 0.19 & 24.379(5) & $-0.15^{+0.04}_{-0.08}$ \nl
 G & $-$1.09 & 0.29 & 24.441(5) & $+0.12^{+0.08}_{-0.03}$ \nl
\multicolumn{5}{c}{}\nl
 H & $-$0.09 & 0.77 & 24.36(3)  & $+0.04\pm0.02$          \nl
 I & $-$1.36 & 0.29 & 24.42(2)  & $+0.15^{+0.15}_{-0.05}$ \nl
 J & $-$0.51 & 1.00 & 24.5867(7)& $+3.4^{+0.9}_{-0.7}$    \nl
 L & $-$2.85 & 0.14 & 24.38(5)  & $+0.3^{+0.3}_{-0.1}$    \nl
 M & $-$1.08 & 1.05 & 24.42(2)  & $+0.12^{+0.06}_{-0.04}$ \nl
\multicolumn{5}{c}{}\nl
 N & $-$0.76 & 0.49 & 24.56(3)  & $+2.9^{+0.7}_{-0.6}$    \nl
 O & $+$1.10 & 0.06 & 24.368(8) & $-0.04\pm0.02$          \nl
 Q & $+$0.80 & 0.98 & 24.29(2)  & $-0.02^{+0.01}_{-0.03}$ \nl
 T & $+$3.84 & 0.34 & 24.39(6)  & $-0.4\pm0.2$            \nl
 U & $+$2.15 & 0.94 & 24.335(4) & $-1.2^{+0.3}_{-0.2}$    \nl
\enddata
\tablecomments{Uncertainties in $(\dot{P}/P)_{\rm obs}$ and
$\theta_\perp$ are less than one in the least-significant digits.  The
values of $(\dot{P}/P)_{\rm obs}$ listed have been obtained from the
actually observed values after a correction for centrifugal and
Galactic acceleration (\cite{fcl+01}).  The uncertainties in the last
quoted digits of the DMs are given in parenthesis. }
\tablenotetext{a}{The line-of-sight distances from the center of
cluster ($R>0$ for the distant half) are model-dependent.  We list
values obtained with the model using a uniform characteristic age,
represented by equation~(\ref{eq:ne2a}).  For the following three
pulsars, the distance or its uncertainty is different when obtained
with the model where $(\dot{P}/P)_{\rm int}$ is drawn from a
distribution (see \S~\ref{sec:ne}): $R_G = +0.2^{+2.4}_{-0.1}$, $R_M =
+0.1^{+1.7}_{-0.1}$, $R_U = -0.14^{+0.05}_{-0.18}$. }
\end{deluxetable}

\end{document}